\documentclass{article}
\usepackage{arxiv}
\usepackage{multirow}
\usepackage{tabularx}
\usepackage{makecell}
\usepackage[utf8]{inputenc} % allow utf-8 input
\usepackage[T1]{fontenc}    % use 8-bit T1 fonts
\usepackage{hyperref}       % hyperlinks
\usepackage{url}            % simple URL typesetting
\usepackage{booktabs}       % professional-quality tables
\usepackage{amsfonts}       % blackboard math symbols
\usepackage{nicefrac}       % compact symbols for 1/2, etc.
\usepackage{microtype}      % microtypography
\usepackage{graphicx}
\usepackage{natbib}
\usepackage{doi}
\usepackage{amsmath}

\title{Where diverse populations gather: Transit accessibility and the spatial structure of social mixing}

\author{ \href{https://orcid.org/0000-0002-6982-1654}{\includegraphics[scale=0.06]{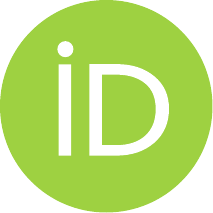}\hspace{1mm}Yuan~Liao}\thanks{Corresponding author. Also affiliated with Department of Applied Mathematics and Computer Science, Technical University of Denmark, Lyngby, Denmark and Department of Space, Earth and Environment, Chalmers University of Technology, Gothenburg, Sweden.} \\
	Department of Human Geography\\
	Lund University\\
	Lund, Sweden \\
	\texttt{yuan.liao@keg.lu.se} \\
}

\renewcommand{\shorttitle}{Spatial Structure of Social Mixing}

\hypersetup{
pdftitle={Where diverse populations gather: Transit accessibility and the spatial structure of social mixing},
pdfsubject={urban mobility, social mixing, transit accessibility},
pdfauthor={Yuan Liao},
pdfkeywords={social mixing, transit accessibility, spatial segregation, points of interest, geographically weighted regression, urban mobility},
}

\begin{document}
\newcommand{\tabincell}[2]{
\begin{tabular}{@{}#1@{}}#2\end{tabular}
}
\maketitle

\begin{abstract}
Urban venues serve as arenas for social mixing, where individuals from different socioeconomic backgrounds share space with one another during daily activities.
While residential and activity-space segregation have been extensively studied, less is known about how the spatial structure of cities, particularly public transit infrastructure, shapes the geography of social mixing at specific locations.
This study examines how transit accessibility associates with visitor diversity --- the compositional heterogeneity of visitors sharing a venue, used here as an indicator of social mixing potential --- at points of interest (POIs) in nine cities in Sweden and three cities in the United States (New York, Washington DC, Atlanta).
Using mobile phone GPS traces from Sweden and aggregated foot traffic data from the US in 2024, we compute visitor diversity indices based on the birth background composition of visitors' home neighborhoods.
We employ spatial regression models and geographically weighted regression (GWR) to test whether transit catchment diversity, the socioeconomic heterogeneity of populations reachable by public transit, predicts visitor diversity, and the spatial heterogeneity of this relationship.
Transit catchment diversity positively predicts visitor diversity across nearly all cities, but this association is robust only in the largest metropolitan areas; in smaller Swedish cities, the coefficient attenuates to insignificance once geographic catchment composition, centrality, and venue density are controlled, indicating that the transit-specific contribution is most detectable where alternative pathways to visitor diversity are limited.
Transit--diversity hotspots tend to occur not in already diverse venues, but in lower-diversity POIs with lower commercial density in both countries' cities, greater distance from transit in the US cities, and greater centrality in Sweden.
These results are consistent with transit infrastructure playing a bridging role in the spatial distribution of visitor diversity, with transit-accessible population composition associated with more diverse co-presence at urban venues.
\end{abstract}

% keywords can be removed

\section{Introduction}

Cities host diverse populations within limited geographic space, yet this proximity does not automatically translate into exposure across group boundaries~\citep{liao2025socio}.
This is partly due to residential segregation, socioeconomic and ethnic groups reside in separate neighborhoods, limiting everyday contact with each other~\citep{burgessResidentialSegregationAmerican1928,masseyAmericanApartheidSegregation1993,musterdSocialEthnicSegregation2005}.
Such everyday contact across group boundaries has only recently become measurable at scale thanks to increasingly available big geolocation data from our mobile phones: a growing body of research now reveals that experienced diversity, the composition of people one actually encounters, often differs substantially from what residential composition alone would predict~\citep{athey2021estimating, moro2021mobility, nilforoshan2023human, renninger2025us, liao2025effect}.
Points of interest (POIs), such as restaurants, shops, and parks, serve as the physical sites where populations from different neighborhoods converge or fail to converge~\citep{oldenburgThirdPlace1982}.
Recent work has demonstrated that certain locations attract visitors from across the urban socioeconomic spectrum, while others primarily serve local, homogeneous populations~\citep{nilforoshan2023human}.
Understanding what makes some places more socioeconomically mixed than others is a central question for urban planning aimed at fostering inclusive cities~\citep{cagneyUrbanMobilityActivity2020}. \par

The diversity of visitors at various urban venues depends on two distinct pools of potential visitors: local visitors who walk in from the surrounding neighborhood, and those who arrive from farther areas via transportation infrastructure~\citep{loo2024residential}.
Visitor diversity at a location thus reflects both the residential composition of its immediate surroundings (residential diversity) and the diversity of populations made accessible by mobility networks.
For example, \cite{moro2021mobility} revealed that 55\% of experienced income segregation in US cities is attributable to mobility patterns rather than residential composition alone.
This decomposition motivates an analytical separation between local residential diversity and transit catchment diversity, the composition of populations reachable through public transport.
This measure captures potential accessibility --- the socioeconomic heterogeneity of populations who could reach a given venue by public transit within a specified travel time --- rather than observed ridership or actual mode choice.
It therefore reflects the diversity of the transit-accessible population, not the composition of visitors who actually arrive by transit.
The key empirical question is the extent to which realized visitor diversity exceeds what the local residential composition alone would predict, and whether that excess is attributable to the populations that transit delivers. \par

Public transit infrastructure plays a demonstrated role in shaping social mixing.
Experienced racial isolation is lower in US metropolitan areas with greater public transit use \citep{athey2021estimating}, bus and metro trips are positively associated with place-level social mixing in Hong Kong \citep{loo2024residential}, and transit structures shape mode-specific income mixing patterns \citep{iyer2024mobility}.
Conversely, limited transit access constrains the mobility of foreign-born minorities, reinforcing their experienced segregation \citep{liao2025effect}.
Yet these findings primarily establish aggregate, city- or individual-level associations between transit provision and segregation outcomes, leaving the spatial mechanisms underspecified.
We propose that transit nodes function as ``spatial bridges'': because transit funnels passengers from neighborhoods of varying composition through common stations and along fixed corridors, the catchment area of a well-connected stop aggregates populations from multiple origins, creating a pool of potential visitors more diverse than any single surrounding neighborhood \citep{nilforoshan2023human}.
This mechanism operates through different channels, including pedestrian dispersal from transit nodes into surrounding commercial areas \citep{hidalgo2020amenity, credit2018transit}, the agglomeration of amenities catering to varied populations near transit hubs
\citep{cervero1997travel, calthorpe1993next}, and trip-chaining behavior that extends the spatial footprint of transit-facilitated visits
\citep{primerano2008defining, ewing2010travel}.
All these channels yield a common prediction: visitor diversity should exhibit spatial spillover, with the diversity of one POI depending on that of nearby locations.
Whether such spillover is structured by the transit network or merely reflects general spatial proximity remains an open question. \par

To examine the spatial structure of social mixing at urban venues, we introduce a cross-national comparison of Swedish and US cities focusing on birth background.
The cities in these two countries differ systematically in urban form, transit orientation, and the structure of social inequality.
The major Swedish cities are compact and transit-oriented, with extensive public transportation networks; segregation there operates along both birth-background and economic lines, with foreign-born minorities disproportionately concentrated in peripheral, lower-income housing estates where daily mobility tends to reproduce rather than reduce residential sorting~\citep{osth2018spatial, hedman2021daily, liao2025effect}.
The US cities span a gradient from transit-rich (New York) to car-dependent, with segregation shaped largely by income and race amid substantially higher inequality~\citep{nilforoshan2023human, moro2021mobility}.
This variation provides analytical leverage.
Let's consider transit accessibility as the transit catchment populations within a certain travel time from a venue.
In the Swedish cities, where transit coverage is extensive and widely used, transit catchment populations may be a weaker differentiator of visitor diversity at venues precisely because most locations are transit-accessible.
In the US cities, where transit access varies sharply across space, transit catchment populations should more strongly distinguish socially mixed venues from homogeneous ones.
We refer to this association between transit catchment diversity and visitor diversity as the \textit{transit--diversity relationship} throughout. \par

Specifically, this study addresses three research questions:
\begin{itemize}
\item To what extent does the socioeconomic diversity of a POI's transit catchment populations predict the realized diversity of its visitors?
\item Does this relationship exhibit spatial spillover effects, and if so, are spillovers better explained by general geographic proximity or by transit-network proximity specifically?
\item How does the transit--diversity relationship vary spatially within cities?
\end{itemize}
Our analysis makes three contributions.
First, we introduce a spatially explicit framework for analyzing social mixing at POIs that distinguishes between residential composition, transit catchment diversity, and realized visitor diversity, connecting the ``supply'' of diversity through transit networks to the mixing actually observed at venues.
Second, we test whether visitor diversity spills over between nearby POIs and, if so, whether this spillover follows the transit network rather than geographic distance alone, providing direct evidence on the patterns linking transit infrastructure to social mixing.
Third, we embed these analyses in a cross-national comparison of the nine Swedish and three US cities, assessing whether the transit--diversity relationship holds across different urban forms and transit systems.
Throughout, we distinguish between the broader concept of social mixing and our empirical measure, visitor diversity, which captures the compositional co-presence of individuals from demographically distinct neighborhoods at the same venue rather than actual social interaction.
Because the analysis is cross-sectional and observational, we interpret results as spatial associations between transit catchment composition and visitor diversity, not as causal effects of transit provision; we return to these boundaries in the Discussion. \par

\section{Results}
\subsection{Analytical overview}\label{sec:overview}
This section presents the results following a three-stage analytical framework, each stage addressing one of the three research questions.
We briefly outline the data, measures, and models here; full details are provided in Materials and Methods (Section~\ref{sec:methods}). \par

The analysis covers approximately 200,000 POIs across three US cities (New York, Washington DC, Atlanta) and 130,000 POIs across nine Swedish cities, using mobile phone-derived mobility data from 2024.
For Sweden, anonymized GPS trajectories from approximately 6.5 million devices are processed into POI visits through stop detection and home-location inference; for the US, we use pre-aggregated weekly foot traffic records reporting visitor counts and home census tract distributions (Sections~\ref{sec:data}--\ref{sec:data_proc}).
It is worth noting that because the Swedish and US analyses rely on different mobility data pipelines, cross-national patterns should be interpreted in terms of directional consistency and qualitative differences rather than direct comparison of effect magnitudes (Section \ref{sec:cross}). \par

Three diversity measures are computed for each POI as normalized entropy (Equation~\ref{eq:entropy}) of visitors' home-neighborhood composition along the birth-background dimension, yielding values from 0 (complete homogeneity) to 1 (even distribution across categories).
\textit{Visitor diversity} ($V$) captures the observed composition of a venue's visitors; \textit{residential diversity} ($R$) reflects the socioeconomic makeup of the census tract containing the POI; and \textit{transit catchment diversity} ($T$) measures the population-weighted heterogeneity of areas reachable within 45 minutes by public transit (Section~\ref{sec:social_mixing}).
A fourth measure, \textit{geographic catchment diversity} ($G$), captures composition within a 1.5~km Euclidean buffer and serves as a robustness control.
Transit catchment diversity reflects potential accessibility --- the diversity of populations who \textit{could} reach a venue by transit --- rather than observed ridership or actual mode choice. \par

The analysis proceeds in three stages.
First, we map and compare the three diversity measures using spatial autocorrelation statistics (Moran's $I$, LISA clusters) to characterize the spatial landscape of visitor diversity (Section~\ref{sec:desc_methods}).
Second, we estimate OLS and spatial lag regression models to quantify the independent association between transit catchment diversity and visitor diversity, testing whether spatial spillovers follow geographic proximity (W1, distance-decayed nearest neighbors) or transit-network connectivity (W2, shared nearest transit stop within 400~m).
Stepwise specifications sequentially add geographic catchment diversity, distance to city center, and local POI density to assess whether the transit catchment coefficient is robust to alternative spatial explanations (Sections~\ref{sec:slm} and surrounding).
Third, we employ geographically weighted regression (GWR) with an adaptive bisquare kernel to examine within-city spatial heterogeneity in the transit--diversity association, classifying POIs as transit--diversity hotspots where the local transit catchment coefficient is both positive and significant under the da~Silva--Fotheringham correction for multiple testing.
A logistic regression then characterizes the locational features --- centrality, commercial density, transit proximity --- that distinguish hotspots from non-significant POIs (Sections~\ref{sec:gwr}--\ref{sec:gwr_hotspots}).

\subsection{Visitor diversity is linked to residential and transit catchment diversity}
\begin{figure}[h!]
\begin{center}
\includegraphics[width=0.8\textwidth]{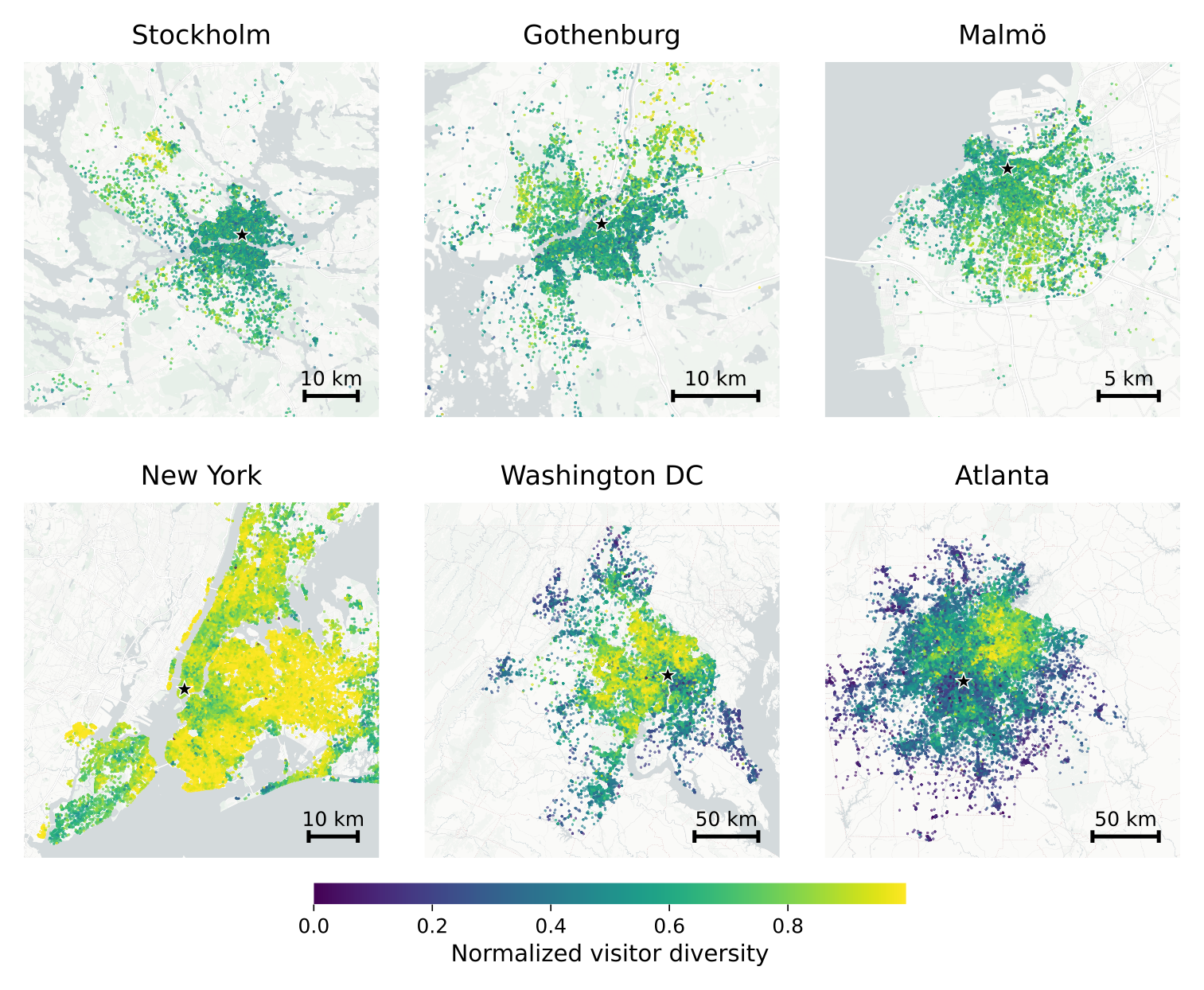}
\end{center}
\caption{\textbf{Birth background diversity of POI visitors across six cities}.
Choropleth maps showing visitor diversity.
Stars mark city centers.}\label{fig:1v}
\end{figure}

We first look at the empirical visitor diversity, which measures the observed birth background composition of actual visitors.
Throughout this paper, we use \textit{visitor diversity} to refer to the compositional heterogeneity of individuals sharing a venue, measured as normalized entropy of visitors' home-neighborhood demographics.
This quantity captures co-presence---the extent to which a venue draws visitors from demographically varied areas---rather than direct social interaction, and we treat it as an indicator of social mixing potential rather than realized social mixing itself.
Focusing on the largest three Swedish cities and the selected US cities, the choropleth maps in Figure~\ref{fig:1v} reveal the spatial organization of visitor diversity by birth background.
Compared to the US cities in the bottom row, the three Swedish cities show lower and more spatially homogeneous visitor diversity. \par

Corroborating what's revealed in Figure~\ref{fig:1v}, Table~\ref{tab:spatial_clustering} quantifies visitor diversity patterns using Moran's I and LISA cluster analysis (full results are available in Supplementary Table S1).
For birth-background diversity, US cities exhibit strong spatial autocorrelation (Moran's I = .67–.83), whereas Swedish cities show much weaker clustering (Moran's I = .13–.24), a difference of roughly three- to sixfold.
The LISA results reinforce this contrast: in US cities, 26–39\% of POIs fall into High-High clusters and 16–29\% into Low-Low clusters, indicating substantial geographic segregation in visitor diversity.
Swedish cities, by comparison, display much more balanced distributions, with only 6--14\% of POIs in each cluster type. \par

\begin{table}[htbp]
  \centering
\caption{Spatial clustering of visitor diversity across Swedish and US cities.
Moran's I measures global spatial autocorrelation; HH\% and LL\% indicate the percentage of POIs in High-High and Low-Low LISA clusters, respectively.
All Moran's I values are significant at $p < .001$.
Results are shown for the three largest Swedish metropolitan areas (Stockholm, Gothenburg, Malmö) alongside the three US cities for compact presentation; full results for all nine Swedish cities are provided in Supplementary Table S1.}
  \label{tab:spatial_clustering}
  \begin{tabular}{llrccc}
  \toprule
  \textbf{Country} & \textbf{City} & \textbf{N POIs} & \textbf{Moran's I} & \textbf{HH\%} &
  \textbf{LL\%} \\
  \midrule
   & Stockholm    & 35,842  & .127 & 6.0  & 5.6  \\
  Sweden & Gothenburg   & 16,685  & .237 & 9.3 & 14.5  \\
   & Malmö        & 12,610  & .197 & 14.3 & 12.8 \\
  \midrule
   & New York       & 148,413 & .833 & 38.6 & 17.7 \\
  US & Atlanta        & 34,354  & .775 & 26.2 & 28.9 \\
   & Washington DC  & 24,599  & .667 & 31.2 & 16.3 \\
  \bottomrule
  \end{tabular}
\end{table}

\begin{figure}[h!]
\begin{center}
\includegraphics[width=0.95\textwidth]{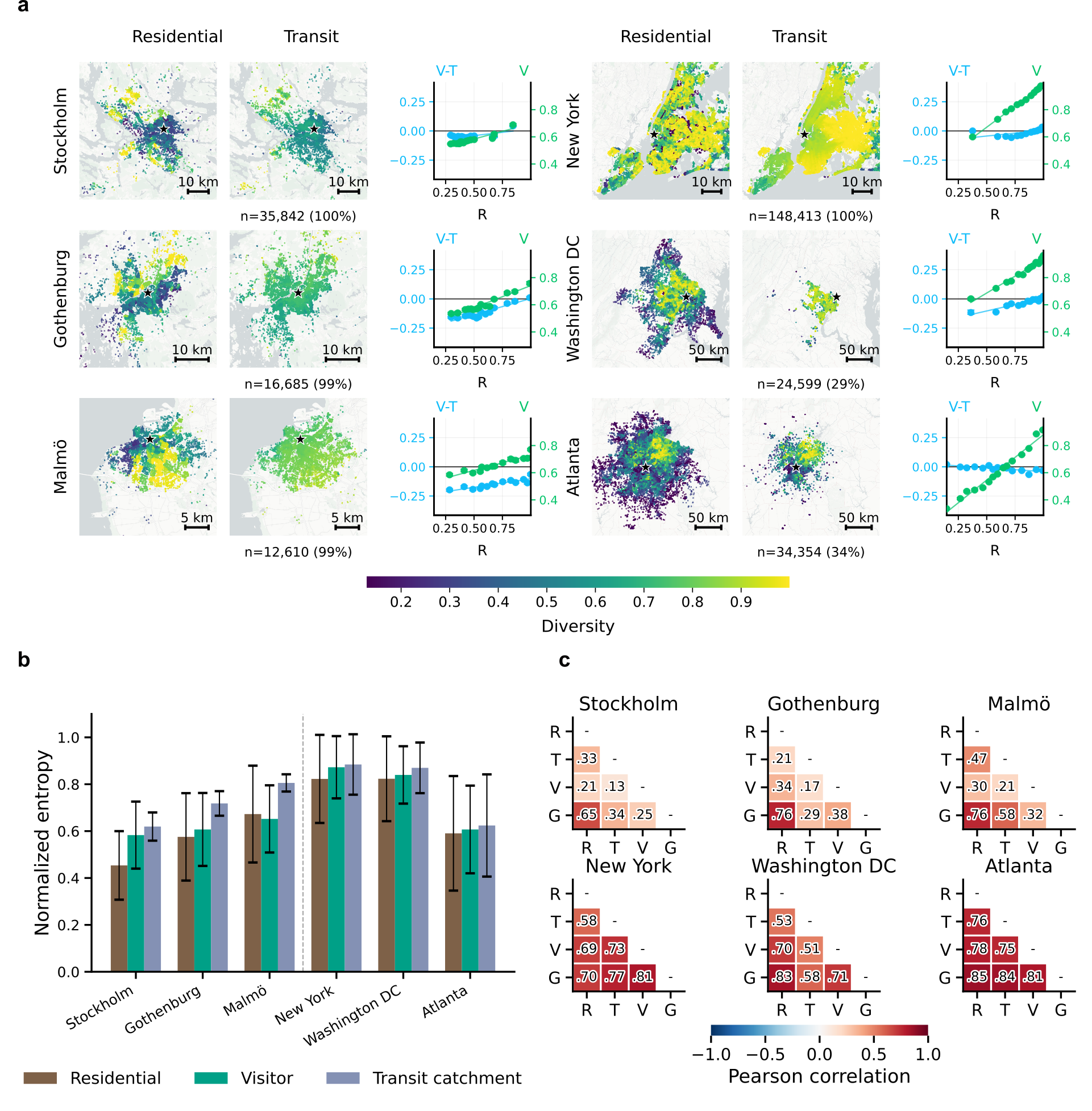}
\end{center}
\caption{\textbf{Visitor diversity links to residential, transit catchment, and geographic catchment diversity at POIs across six cities.}
(\textbf{A}) Spatial distribution of normalized birth-background diversity for POIs in three Swedish cities and three US cities.
For each city, the left map shows residential diversity at the POI location, and the right map shows transit catchment diversity for POIs with a valid transit catchment.
Stars indicate city centers.
Values below each city pair report the number and percentage of POIs with transit catchment information.
Inset plots summarize binned (15 percentiles) POI-level relationships with residential diversity ($R$): visitor diversity ($V$, green) and the difference between visitor and transit catchment diversity ($V-T$, blue).
(\textbf{B}) Distribution of normalized entropy by city for POIs with a transit catchment.
Bars show median diversity for residential populations, visitors, and transit catchment populations; error bars indicate standard deviations.
(\textbf{C}) Lower-triangle Pearson correlation matrices comparing residential diversity ($R$), transit catchment diversity ($T$), visitor diversity ($V$), and geographic catchment diversity ($G$) at the POI level.}
\label{fig:1r}
\end{figure}

We then examine how empirical visitor diversity is linked to the other two types of diversity measures at POIs (Figure~\ref{fig:1r}): residential diversity ($R$), which captures the socioeconomic composition of the census tract containing the POI; transit catchment diversity ($T$), which reflects the composition of areas reachable within 45 minutes by public transit.
In addition, we consider geographic catchment diversity within 1.5 km ($G$), reflecting the birth background composition of neighborhoods within walking distance.

Across all six cities, residential diversity exhibits stronger spatial clustering than transit catchment diversity, particularly in the three Swedish cities (Figure~\ref{fig:1r}a, choropleth maps).
In the Swedish cities, transit catchment diversity appears to operate as a spatial smoothing mechanism, linking otherwise residentially segregated areas.
While in the three US cities, the transit catchment diversity still presents clear spatial heterogeneity.
Consistent with the choropleth maps (Figure~\ref{fig:1r}a), both visitor diversity and its deviation from transit catchment diversity generally scale with residential diversity (Figure~\ref{fig:1r}a, scatter plots).
Except for Atlanta, all other cities show a clear association between the visitor--catchment diversity gap and residential diversity.
It is worth noting that, in Washington DC and Atlanta, not all POIs have corresponding transit catchment diversity (see Figure \ref{fig:1r}a, numbers under the Transit columns) due to a lack of census tract coverage within the travel time threshold of 45 min by public transit (inferred from GTFS data).
Washington DC and Atlanta span large geographical areas, while transit services are concentrated primarily in their urban cores.
To further examine how visitor diversity relates to other dimensions of social mixing, we focus on POIs for which all social-mixing measures are available (Figures~\ref{fig:1r}b–c).
The Swedish cities show lower median visitor diversity than the US cities (.51--.81 vs. .63--.89).
Residential, visitor, and catchment diversity differ significantly (pairwise Wilcoxon signed-rank tests $p < .001$, with a magnitude of .02--.2), confirming that the three measures capture different aspects of POI-level diversity (see Supplementary Tables~S2). \par

The patterns of social mixing differ markedly, regarding how visitor diversity is linked with residential and transit catchment diversity by country (Figures \ref{fig:1r}b and Supplementary Table~S2).
Both studied Swedish and US cities generally show higher visitor diversity than residential diversity, but the pattern is modest and uneven.
The largest gains appear in Stockholm (+.129), while several Swedish cities show lower visitor than residential diversity, including Malmö (-.021), Helsingborg (-.015), and Linköping (-.008).
Transit catchment diversity is consistently higher than residential diversity across all cities, indicating that transit-accessible catchments are broader and more diverse than both local residential contexts and observed visitor populations. \par

Taken together, Figure~\ref{fig:1r}c shows that the different measures of social mixing are much more tightly coupled in US cities than in Swedish cities.
In Stockholm, Gothenburg, and Malmö, visitor diversity is only weakly to moderately correlated with residential diversity ($r=.21$--$.34$) and even more weakly with transit catchment diversity ($r=.13$--$.21$), suggesting that POI-level mixing is not simply a reflection of nearby residential composition or transit-reachable populations.
By contrast, in New York, Washington DC, and Atlanta, visitor diversity is strongly correlated with residential diversity ($r=.69$--$.78$), transit catchment diversity ($r=.51$--$.75$), and geographic catchment diversity ($r=.71$--$.81$).
In summary, visitor diversity at POIs is spatially diffuse in Swedish cities but strongly clustered in US cities, where it closely tracks residential segregation patterns.

\subsection{Transit catchment diversity predicts visitor diversity in big cities with spatial spillovers driven by proximity}
Having established the spatial landscape of visitor diversity, we next ask whether transit catchment diversity predicts visitor diversity and whether visitor diversity spills over between neighboring POIs.
We estimate OLS regressions to quantify the association between transit catchment diversity and visitor diversity, controlling for POI category.
The robustness of this effect is tested through stepwise OLS specifications that sequentially add geographic catchment diversity, distance to the city center, and POI density within 500 m, allowing us to assess whether the transit catchment coefficient remains stable after accounting for local composition, spatial proximity, centrality, and venue agglomeration. \par

\begin{figure}[h!]
\begin{center}
\includegraphics[width=0.9\textwidth]{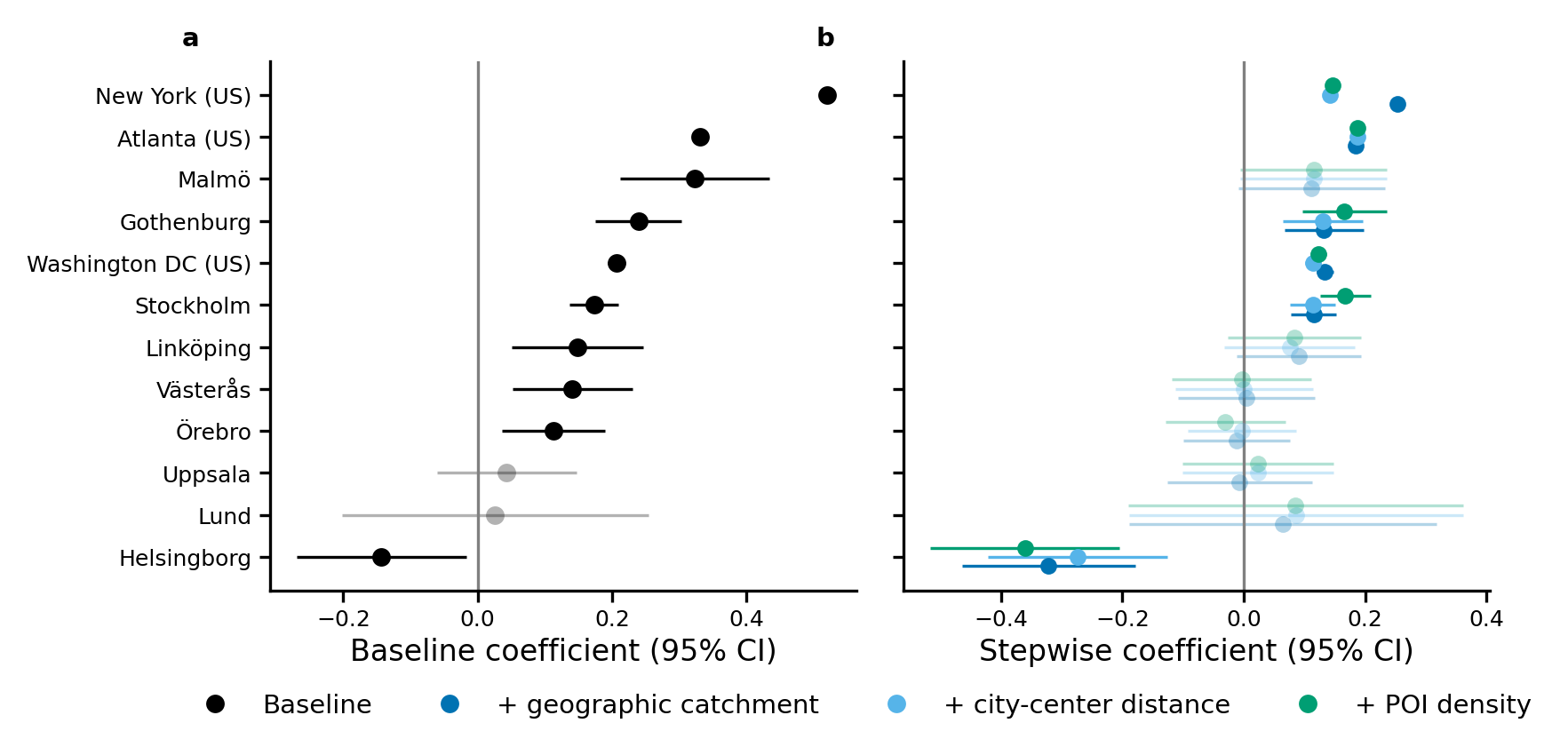}
\end{center}\caption{\textbf{Transit catchment diversity as a predictor of visitor diversity across cities}.
(\textbf{A}) OLS regression coefficients for transit catchment diversity predicting POI visitor diversity.
Cities are ranked from largest to smallest coefficient.
Points indicate point estimates; horizontal lines show 95\% confidence intervals.
Faded points with confidence intervals crossing zero indicate non-significant coefficients.
US cities are labeled accordingly; all others are Swedish cities.
(\textbf{B}) Stepwise robustness checks for the birth-background models.
Points show the transit catchment diversity coefficient after sequentially adding geographic catchment diversity, distance to the city center, and POI density within 500 m; horizontal lines indicate 95\% confidence intervals.}\label{fig:2}
\end{figure}

The transit catchment diversity coefficient is positive and significant in nearly all cities, confirming that POIs whose transit catchments draw from more diverse populations also tend to attract more diverse visitors (Figure~\ref{fig:2}).
First, we find that the magnitude of this relationship varies substantially across cities (Figure~\ref{fig:2}a).
The three US cities and the three largest Swedish cities exhibit the largest coefficients, with New York standing out at approximately .5.
Atlanta and Malmö follow closely, around .3.
Gothenburg and Washington DC cluster in the .2 range.
The mid-sized Swedish cities, Uppsala, Linköping, Helsingborg, Västerås, and Örebro, show smaller but still significant coefficients between .1 and .2.
Second, when stepwise robustness checks add geographic catchment diversity, distance to the city center, and POI density within 500 m, the transit catchment coefficient generally attenuates but remains positive and statistically significant in the US cities and the largest Swedish cities (Figure~\ref{fig:2}b).
In the smaller Swedish cities, however, the coefficient becomes statistically insignificant, suggesting that the apparent transit catchment effect is largely absorbed by broader spatial structure.
Helsingborg is the clearest counterexample.
Its coefficient is already negative in the baseline model and becomes even more negative after spatial controls are added, suggesting that transit catchment diversity does not uniformly translate into observed visitor diversity in smaller cities.
One plausible interpretation is that, in Helsingborg, the diversity reachable by transit is largely absorbed by broader residential and geographic catchment structure, while the residual transit-specific variation may reflect peripheral or functionally specialized transit-accessible locations that attract less diverse visitor flows.
We therefore interpret this result cautiously as evidence that transit catchment diversity is not a universal proxy for realized visitor diversity, especially in smaller urban systems. \par

Then, we analyze the spatial spillovers of visitor diversity, understanding whether such spatial interdependence is stronger under ordinary geographic proximity or transit-structured proximity.
To test for spatial spillovers, we estimate spatial lag models that add a spatially weighted average of neighboring POIs' visitor diversity, $\rho \sum_j w_{ij} V_j$, under two alternative weight specifications: W1 connects each POI to its nearest neighbors in a distance-decayed manner, capturing general spatial clustering; W2 connects POIs sharing the same nearest transit stop within 400 meters, capturing transit-mediated connectivity specifically (see Section~\ref{sec:slm} for model details and weight construction).
If transit infrastructure channels diversity spillovers between POIs, we would expect W2 to yield a larger spatial autoregressive coefficient $\rho$ than W1.
Models are estimated separately for each city.
When considering the neighboring POIs, the spatial lag models outperform the OLS baselines in predicting visitor diversity (Table \ref{tab:model_performance_birth}). \par

\begin{table}[htbp]
  \centering
  \caption{Model performance comparison between OLS (baseline) and spatial lag models.
  W1 and W2 denote the two spatial weight specifications; $\rho$ is the spatial autoregressive coefficient.
  $^{***}p<.001$, $^{**}p<.01$, $^{*}p<.05$.
  N (K) indicates sample size in thousands.}
  \label{tab:model_performance_birth}
\begin{tabular}{lrrrrrr}
\toprule
 & & & \multicolumn{2}{c}{\textbf{W1}} & \multicolumn{2}{c}{\textbf{W2}} \\
\textbf{City} & \textbf{N (K)} & \textbf{OLS $R^2$} & \textbf{$\rho$} & \textbf{$R^2$} & \textbf{$\rho$} & \textbf{$R^2$} \\
\midrule
Gothenburg    & 7.7   & .120 & .848$^{***}$ & .193 & --              & --   \\
Helsingborg   & 2.4   & .070 & .891$^{***}$ & .099 & --              & --   \\
Linköping     & 2.3   & .052 & .908$^{***}$ & .089 & --              & --   \\
Lund          & 2.0   & .015 & .563$^{*}$   & .020 & --              & --   \\
Malmö         & 5.6   & .079 & .891$^{***}$ & .120 & --              & --   \\
Stockholm     & 16.2  & .054 & .836$^{***}$ & .084 & -.027$^{**}$    & .053 \\
Uppsala       & 3.3   & .047 & .965$^{***}$ & .073 & -.005           & .046 \\
Västerås      & 2.2   & .034 & .723$^{**}$  & .060 & --              & --   \\
Örebro        & 2.4   & .039 & .804$^{***}$ & .056 & --              & --   \\
\midrule
Atlanta       & 28.5  & .674 & .514$^{***}$ & .768 & -.011$^{***}$   & .673 \\
New York      & 127.6 & .641 & .769$^{***}$ & .813 &  .039$^{***}$   & .658 \\
Washington DC & 19.6  & .514 & .497$^{***}$ & .663 &  .025$^{***}$   & .525 \\
\bottomrule
\end{tabular}
\end{table}

On birth-background social mixing, we see strong spatial spillover effects of visitor diversity and a systematic difference between the studied Swedish and US cities in both baseline model fit and the nature of spatial dependence.
The spatial lag models with W1, which connects each POI to its nearest neighbors (with distance decay), consistently outperform OLS, with $\rho$ positive and significant in all cities.
The coefficient $\rho$ under W1 ranges from approximately .50 in Washington DC to .97 in Uppsala, meaning that when nearby POIs attract diverse visitors, a given POI tends to do the same.
However, W2, which connects POIs that share the same nearest transit stop, yields $\rho$ values that are either non-significant or negligibly small in the Swedish cities where W2 could be estimated.
In Stockholm, W2 is small and negative ($\rho=-.027$), while in Uppsala it is effectively zero and non-significant ($\rho=-.005$).
In other words, knowing that two venues sit near the same bus or train stop does not help predict whether they share similar levels of visitor diversity in the Swedish cases studied.
This suggests that in the studied Swedish cities, the spatial pattern of social mixing reflects broader neighborhood characteristics rather than a transit-station-specific spillover mechanism. \par

The US cities tell a different story in two respects.
First, the basic OLS models already explain far more of the variation in birth-background diversity, with $R^2$ values of .51-.67, compared with .02-.12 for Swedish cities.
This likely reflects the stronger spatial structure of racial and ethnic segregation in US metropolitan areas, where residential composition and transit geography are more tightly coupled.
Second, W2 produces statistically significant spatial dependence across all three US cities, although the direction and magnitude of this dependence vary.
New York and Washington DC exhibit a positive coefficient, indicating that transit-mediated connectivity matters: venues clustered around the same station tend to share similar diversity levels under the W2 specification.
However, Atlanta shows a small but significant negative coefficient, indicating a tendency toward dissimilarity among POIs sharing transit connections.
Still, W1 delivers larger improvements in model fit than W2, suggesting that geographic clustering remains the dominant spatial pattern even where transit spillovers are detectable. \par

Together, we identify that transit catchment diversity is a significant predictor of visitor mixing in big cities, with the strongest effects in large cities like New York and in cities where transit networks bridge demographically distinct neighborhoods, such as Gothenburg and Atlanta.
Spatial lag models confirm that diverse venues cluster near other diverse venues, but this clustering follows general geographic proximity (W1) rather than shared transit stops (W2) in Swedish cities.
Only in the US cities does transit-specific connectivity contribute significant additional explanatory power.
These findings are consistent with the interpretation that transit infrastructure is more closely associated with the diversity of populations reachable from surrounding areas, as captured by the catchment diversity measure, than with direct spillovers between venues near the same stop.
The role of transit in social mixing may therefore be indirect: it appears to be linked to the composition of accessible nearby populations, while neighborhood-level spatial structure helps account for how diversity is distributed across local venues.

\subsection{Spatial variation in transit effects on birth-background visitor diversity}
The preceding analyses estimate global relationships that assume the transit--diversity association is spatially constant within each city.
To examine whether this relationship varies across space, and to identify where transit catchment diversity most strongly translates into realized visitor mixing, we employ geographically weighted regression (GWR).
GWR estimates location-specific coefficients for each POI using an adaptive bisquare kernel, with bandwidth optimized by corrected AIC (see Section~\ref{sec:gwr}).
Statistical significance of local coefficients is assessed using the da~Silva--Fotheringham correction for multiple testing.
We classify POIs as \textit{transit--diversity hotspots} where the local transit catchment coefficient is both positive and significant, then characterize these hotspots by their distance to the city center, surrounding POI density, and proximity to the nearest transit stop.
A logistic regression jointly models the predictors of hotspot status (see Section~\ref{sec:gwr_hotspots}). \par

Across cities, the GWR results (all POIs by city and social-mixing dimension) reveal substantial spatial heterogeneity in the relationship between transit catchment diversity and visitor diversity (Supplementary Table S3).
Most studied cities (8 out of 12) exhibit positive local coefficients, indicating a generally positive association between transit accessibility and birth-background visitor diversity, although the magnitude and spatial concentration of this effect vary markedly.
Notably, the three US cities, show much stronger and more spatially concentrated effects, reflected in high hotspot shares (7.5--32.5\%) and substantially higher model fit ($R^2 \approx .70-.80$), whereas Swedish cities display weaker and more diffuse patterns with lower $R^2$ values. \par

\begin{figure}[h!]
\begin{center}
\includegraphics[width=0.8\textwidth]{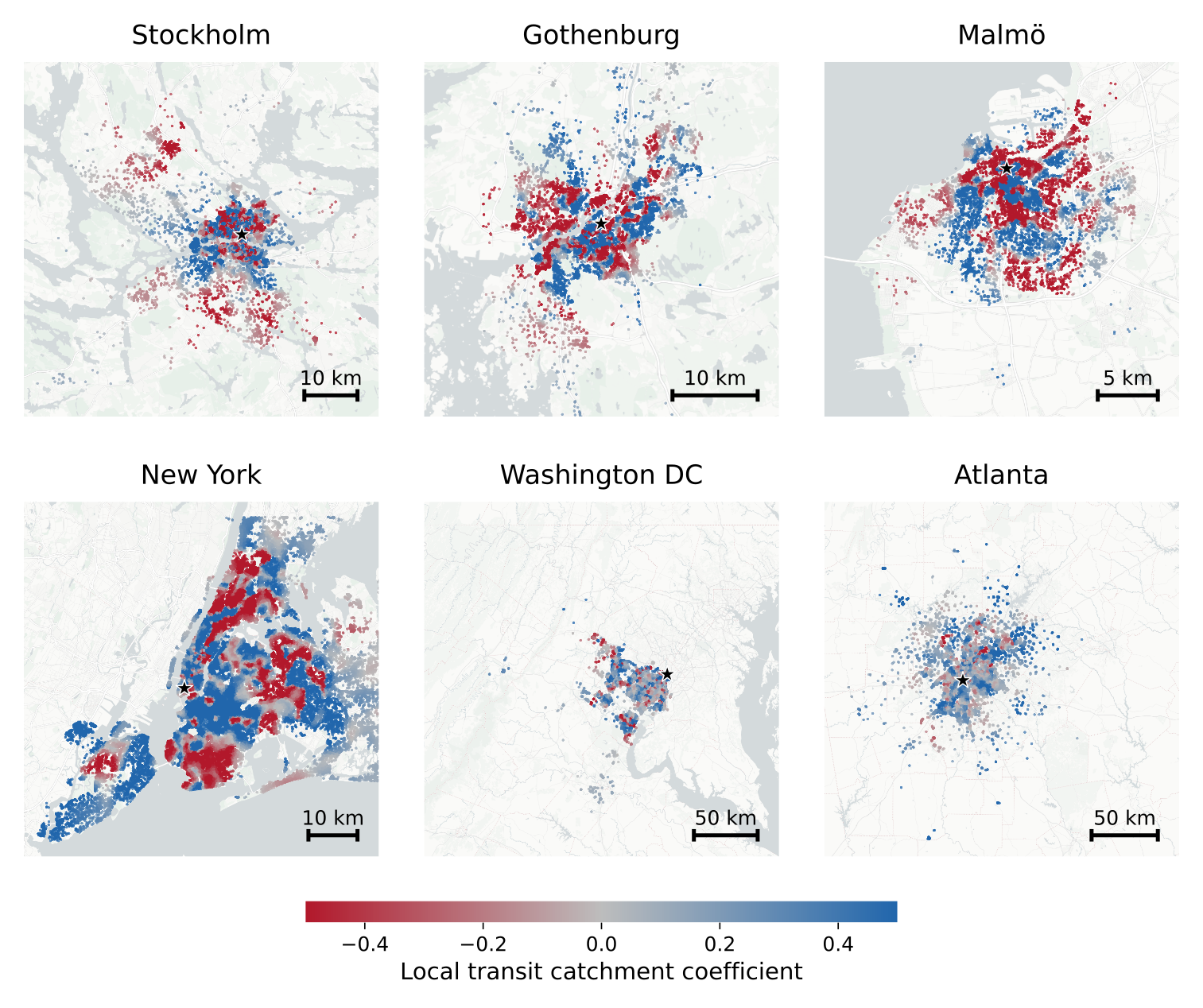}
\end{center}\caption{\textbf{Spatial distribution of local coefficients from geographically weighted regression (GWR) models estimating the relationship between transit catchment diversity and visitor diversity (all POIs)}.
Each point represents a POI, colored by the estimated local coefficient, with blue indicating positive associations and red indicating negative associations; gray denotes near-zero effects.
Results are shown for the three Swedish cities (top row) and the three US cities (bottom row). }\label{fig:3}
\end{figure}

The spatial distribution of local coefficients reveals striking cross-national differences in the transit--diversity relationship.
In the three US cities, the association between transit catchment diversity and visitor diversity is predominantly positive and spatially consistent (Figure \ref{fig:3}), particularly in New York.
In contrast, Swedish cities exhibit weaker, more spatially heterogeneous patterns.
Notably, only 0-8.8\% of Swedish POIs qualify as significant hotspots after correcting for multiple testing (Supplementary Table S3), suggesting that the transit--diversity relationship is not statistically distinguishable from zero at most individual locations.
This pattern is consistent with Sweden's more uniform visitor diversity, which limits the potential for transit to generate spatial diversity gradients.
We therefore interpret these hotspots as locations where transit catchment diversity is a statistically significant positive predictor of visitor diversity, rather than as direct evidence that transit delivers diverse populations to specific POIs. \par

\begin{figure}[h!]
\begin{center}
\includegraphics[width=0.95\textwidth]{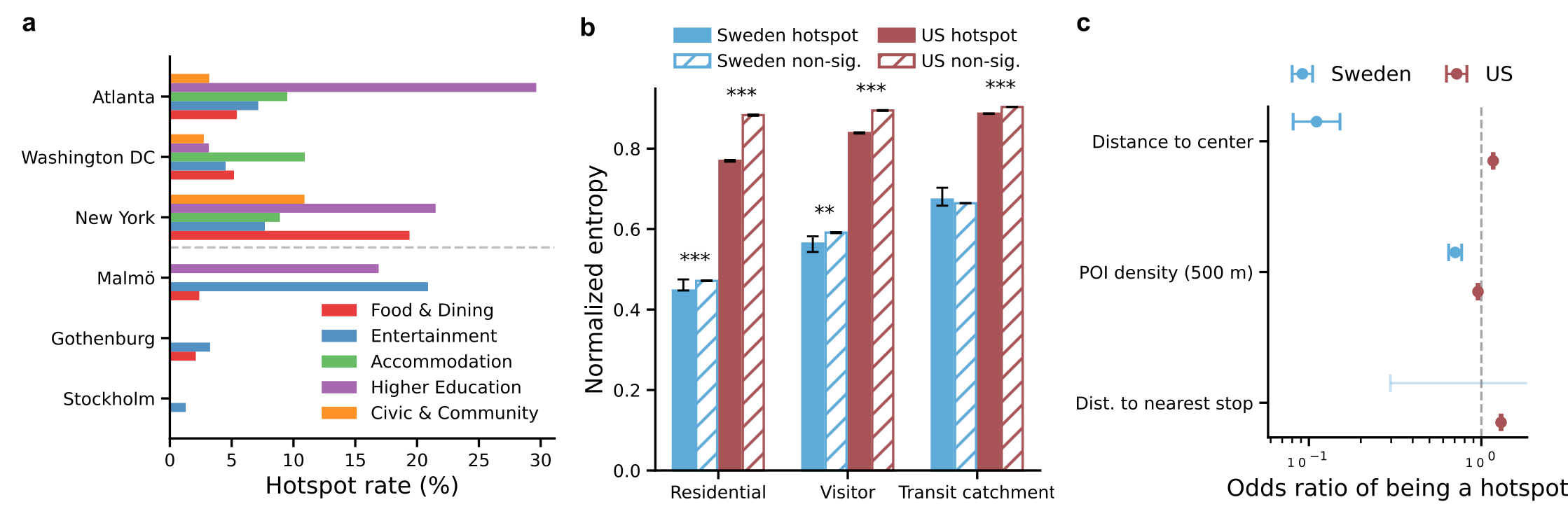}
\end{center}\caption{\textbf{Characteristics of transit catchment diversity hotspots across cities}.
(\textbf{A}) Hotspot rates by POI category, showing the percentage of locations where transit catchment diversity has a significant positive association with visitor diversity.
(\textbf{B}) Diversity characteristics of transit--diversity hotspots by country.
Bars show median normalized entropy, comparing hotspot POIs with non-significant POIs pooled separately for studied cities in Sweden and the US.
Error bars indicate bootstrapped 95\% confidence intervals for the median.
Asterisks mark significant hotspot versus non-significant differences within the country based on Mann-Whitney U tests.
(\textbf{C}) Odds ratios from logistic regression models predicting hotspot status based on distance to city center, local POI density (within 500 m), and distance to the nearest transit stop, estimated separately for the Swedish and the US cities studied.
Points indicate odds ratios; horizontal bars indicate 95\% confidence intervals. The dashed vertical line marks an odds ratio of 1 (no effect).
}\label{fig:4}
\end{figure}

The spatial heterogeneity in the transit catchment diversity effect raises a natural question: where within cities does transit catchment diversity most strongly predict realized visitor diversity?
Transit--diversity hotspots --- locations where transit catchment diversity significantly predicts visitor diversity --- exhibit distinct spatial and contextual characteristics that differ between the studied Swedish and US cities (Figure \ref{fig:4}).
Hotspot rates vary substantially by venue category and urban context, with US cities showing higher overall rates, particularly for food and dining establishments (Figure \ref{fig:4}a).
Compared with non-significant POIs, hotspots show a consistent profile: they tend to occur in areas with lower residential, visitor, and transit diversity (Figure~\ref{fig:4}b). \par

In both countries, hotspots are located in less-dense commercial clusters (Figure \ref{fig:4}c, Supplementary Table S4).
But they diverge on distance to the city center and transit proximity.
In US cities, hotspots tend to concentrate in peripheral areas where the surrounding residential population is relatively homogeneous.
The greater distance to transit stops further implies that these are not immediate station-area effects but rather diffuse spillovers along corridors where transit-connected visitors disperse into neighborhoods with few alternative sources of population mixing.
In the Swedish cities studied, by contrast, hotspots emerge in places closer to the city center, indicating that transit amplifies the pre-existing attractiveness of the city center to diverse populations rather than compensating for its absence.
Here, the patterns agree with the reinforcement mechanism: attractive central locations combine with diverse transit-delivered flows.
The shared pattern --- locations in less-dense commercial clusters --- suggests that, in both contexts, central, high-density commercial areas may already achieve visitor diversity through sheer volume and variety of visitors, regardless of transit catchment composition.
The hotspots' tendency to occur in more peripheral and lower-density contexts is consistent with a transit-bridging mechanism.
This aligns with the theoretical expectation that transit's role in shaping social mixing should be most detectable where alternative pathways to diversity are limited. \par

In summary, while hotspot formation in both urban systems is negatively associated with dense local commercial environments, it is not concentrated in immediately transit-adjacent locations.
Instead, hotspots appear to arise in more spatially selective settings, particularly in the studied US cities, while in the 9 cities of Sweden, transit seems to reinforce the most central locations.

\section{Discussion}

This study examined how transit accessibility predicts the spatial structure of social mixing at urban venues across nine Swedish and three US cities.
By distinguishing between residential diversity, transit catchment diversity, and realized visitor diversity, we tested whether the socioeconomic heterogeneity of populations reachable by public transit predicts the diversity of visitors at points of interest, whether this relationship exhibits spatial spillovers, and how the transit--diversity association varies across urban space.
Our findings reveal consistent but context-dependent relationships regarding the potential and limits of transit infrastructure as a facilitator of social mixing. \par

Our results indicate that transit catchment diversity is positively associated with visitor diversity at POIs across nearly all cities, but that this relationship is robust primarily in the largest metropolitan areas.
After accounting for local residential composition, geographic catchment diversity, centrality, and nearby venue density, the transit catchment coefficient remains significant in large cities but attenuates to insignificance in smaller Swedish cities.
We therefore interpret transit catchment diversity not as direct evidence of a causal mechanism, but as an indicator of the spatial opportunity structure through which transit may expand the range of socially diverse populations able to access a venue.
This interpretation is consistent with prior work showing that mobility patterns contribute to experienced segregation beyond residential sorting alone~\citep{moro2021mobility, athey2021estimating}.
Our contribution is to show that the composition of populations reachable by transit is systematically related to venue-level diversity, especially in large metropolitan contexts where transit infrastructure more clearly structures access across otherwise separated neighborhoods.
In smaller and less spatially differentiated cities, by contrast, visitor diversity appears to be explained more fully by broader geographic accessibility and urban centrality, suggesting that transit-specific opportunity structures are harder to distinguish from the general urban fabric. \par

Despite revealing that transit is a robust predictor of visitor mixing levels in big cities, the spatial lag model results complicate a simple ``transit as connector'' claim.
We hypothesized that if transit nodes function as spatial bridges, visitor diversity should spill over more strongly between POIs sharing the same transit stop (W2) than between POIs that are merely geographically proximate (W1).
This prediction was NOT supported in the studied Swedish cities, where W2 contributed no additional explanatory power beyond general proximity.
In the studied US cities, W2 showed significant spatial dependence, but W1 consistently yielded larger improvements in model fit.
This pattern suggests that transit shapes visitor diversity primarily by channeling diverse populations into surrounding areas --- as captured by the catchment diversity measure --- rather than through direct station-to-station spillovers.
The revealed patterns seem to be consistent with a indirect mechanism: transit delivers diverse populations to neighborhoods, where local processes, such as walking to nearby venues or visiting multiple destinations in a single trip, then distribute that diversity across proximate POIs. \par

A central contribution of this study is the identification of distinct spatial patterns linking transit to visitor diversity across the two national contexts.
In the studied US cities, transit--diversity hotspots concentrate in lower-diversity, peripheral locations where the surrounding residential population is relatively homogeneous, suggesting that transit functions as a \textit{diversity bridge}, importing heterogeneous visitor flows into areas that would otherwise attract uniform populations.
In the studied Swedish cities, by contrast, hotspots emerge in already-dense and central neighborhoods closer to the city center, indicating that transit \textit{amplifies} pre-existing central attractive locations.
This divergence likely reflects differences in urban form and segregation structure: studied Swedish cities exhibit more spatially diffuse visitor diversity than the strong clustering observed in the three US cities, limiting the potential for transit to generate diversity gradients as sharp as those in US cities. \par

The shared finding across both national contexts --- that hotspots tend to occur in areas with relatively low residential and visitor diversity and lower commercial density --- offers additional insight.
Commercially dense areas (high POI density) may already achieve visitor diversity through agglomeration and spillover effects, regardless of transit catchment composition.
It is in the area where visitor pools are thinner and more dependent on specific access channels that transit catchment composition becomes a decisive predictor.
This aligns with theoretical expectations that transit's role in shaping social mixing should be most detectable where alternative pathways to diversity are limited~\citep{nilforoshan2023human}. \par

Our findings carry suggestive implications for urban planning and transit policy aimed at fostering inclusive cities, though we emphasize that the associational nature of our evidence means these should be treated as hypotheses for further investigation rather than established prescriptions.
The positive association between transit catchment diversity and visitor diversity is consistent with the idea that expanding transit access to demographically diverse catchments may be associated with social mixing at destination venues, but causal, longitudinal, or quasi-experimental evidence would be needed to establish this link directly.
The cross-national differences further caution against one-size-fits-all prescriptions.
In contexts resembling US cities, where residential segregation is spatially concentrated and transit coverage is uneven, our results suggest that extending transit to peripheral, lower-diversity areas may have the greatest potential to introduce diverse visitor flows where they are otherwise absent.
Transit-oriented development strategies that prioritize connectivity across demographically distinct neighborhoods, rather than simply densifying already-accessible central areas, may be most effective at leveraging this bridging potential, though this remains to be tested with designs capable of identifying causal effects.
In contexts resembling Swedish cities, where transit coverage is already extensive and residential diversity more spatially diffuse, the results point toward a different policy lever: ensuring that transit-served areas maintain diverse residential populations.
Here, housing policies that prevent socioeconomic homogenization of transit-accessible neighborhoods may be as important as transit investments themselves.
In both contexts, the finding that central, high-density commercial areas do not exhibit stronger transit--diversity associations suggests that the mixing correlates of transit are not automatic consequences of agglomeration.
Planners may wish to consider the full spatial footprint of transit networks, including peripheral nodes, rather than focusing solely on central hubs.
Further research using quasi-experimental or longitudinal designs is needed to determine whether these spatial patterns reflect causal relationships that policy interventions could leverage. \par

\subsection{Limitations and future work}
This study has several limitations.
First, birth-background measures are not directly comparable across countries, as Sweden distinguishes native- vs.\ foreign-born (excluding Europe-born) while the US captures racial and ethnic composition; results should be interpreted within rather than across national contexts.
Second, the analysis is observational, and although we control for key factors and account for spatial dependence, unobserved confounders may remain.
Third, mobile phone data may be subject to selection bias despite weighting and large sample sizes.
Smartphone ownership and app usage vary by age, income, and immigration status \citep{cabrera2025systematic}, and the two countries rely on different data pipelines --- opt-in GPS tracking in Sweden versus aggregated app-based foot traffic in the US --- each with distinct selection profiles and home-location detection methods.
While inverse probability weighting partially corrects for known demographic imbalances in the Swedish data \citep{liao2025effect}, residual selection effects likely remain in both contexts and may differ in direction across countries.
Fourth, key parameters, such as the 45-minute catchment threshold, are somewhat arbitrary, though the results appear robust.
Fifth, our transit catchment diversity measure captures the composition of populations \textit{potentially} reachable by public transit, not the mode of transport visitors actually used.
This distinction has asymmetric implications across contexts: in car-oriented settings such as Atlanta, where many visitors likely arrive by car, transit catchment diversity may overstate the role that public transit actually plays in assembling diverse visitor pools; conversely, in transit-rich settings, where most visitors may arrive by transit regardless of catchment composition, the measure may understate transit's contribution by failing to distinguish transit-facilitated visits from the broader population pool.
Finally, we measure co-presence rather than actual social interaction, meaning observed diversity reflects the potential for contact rather than realized visitor diversity. \par

Future work could strengthen causal identification, test alternative definitions of accessibility, improve cross-national harmonization of diversity measures, and link co-presence more directly to realized social interaction and longer-term social outcomes.
There are several extensions that could strengthen the identification of a transit-specific effect.
Neighborhood fixed effects would help absorb unobserved tract-level confounders (land-use mix, institutional presence).
Employment density controls would address the possibility that job-center POIs attract diverse visitors through commuter flows rather than transit access.
Panel data tracking visitor diversity before and after transit network changes, richer land-use covariates, and larger multi-city pooled designs would enable these analyses.
Additionally, one could further assess the sensitivity of hotspot classification by using local coefficients from MGWR as an alternative basis for defining transit--diversity hotspots, and by testing whether the spatial profiles identified by the logistic regression remain stable across different hotspot definitions, catchment thresholds, and model specifications.
While this study focuses on birth-background diversity, preliminary analyses suggest that income-based visitor mixing is less spatially structured and less well captured by the transit catchment framework, likely reflecting the stronger role of destination pricing, amenity type, and temporal sorting in shaping income-based mixing.
Extending the analytical framework to income and other dimensions of social stratification -- such as education and age -- remains an important direction for future work.

\subsection{Conclusions}
In summary, this study provides the first spatially explicit, cross-national analysis of how transit accessibility predicts social mixing at urban venues.
We find that transit catchment diversity consistently predicts visitor diversity in the studied big cities, but the spatial structure of this relationship differs markedly across the nine studied Swedish and US cities (New York, Washington DC, and Atlanta).
In the three US cities, transit appears to function as a diversity bridge, importing heterogeneous visitor flows into otherwise homogeneous peripheral areas.
In the nine Swedish cities, transit seems to amplify pre-existing attractive venues in central cities.
Spatial spillovers in visitor diversity are associated primarily with geographic proximity rather than transit-network connectivity, suggesting that transit's role in shaping visitor composition operates through the diversity of populations reachable within catchment areas rather than through direct station-level linkages between venues.
These findings advance our understanding of how urban infrastructure mediates social stratification in everyday life.
They also highlight the importance of context: the same infrastructure may operate through different mechanisms depending on the urban form, segregation structure, and transit coverage in which it is embedded.
As cities worldwide invest in public transit to address climate, equity, and livability goals, understanding these contingent relationships is essential for designing systems that foster not only mobility but also the social mixing that proximity makes possible.

\section{Materials and Methods}\label{sec:methods}

\subsection{Study Areas}\label{sec:study_areas}
We examine nine cities in Sweden and three cities in the United States, representing distinct urban mobility regimes.
The Swedish cities include Stockholm, Gothenburg, and Malmö, the country's three largest metropolitan areas, along with six mid-sized cities (Uppsala, Västerås, Örebro, Linköping, Helsingborg, and Lund).
These cities are characterized by relatively integrated public transit networks and compact urban forms.
The US cities, New York, Washington DC, and Atlanta, span a range of transit provision, from New York's dense subway network to Washington DC's radial metro system to Atlanta's car-dependent metropolitan form.
In addition, the selected US cities are substantially larger in population (6–19 million) than the Swedish cities (.7–2.4 million).

\subsection{Data Sources}\label{sec:data}
\subsubsection{Points of Interest}
Venues of social mixing in this study are operationalized using points of interest (POIs).
For Sweden, POI data were obtained from SafeGraph Global Places \citep{safegraph2022globalplaces}.
The US mobility data come with integrated POI information from the same underlying source.
The dataset provides comprehensive coverage of physical locations where people spend time or money, including restaurants, retail stores, recreational facilities, and public services.
Each POI record includes geographic coordinates, address, brand affiliation, and category classification using NAICS codes.

\subsubsection{Mobility Data}
For Sweden, we use anonymized GPS trajectory data collected from approximately 6.5 million mobile devices throughout 2024.
The raw geolocation records are processed using Infostop~\cite{aslak2020infostop}, a stop-detection algorithm to identify stationary periods that may represent potential POI visits.
Stays are defined as sequences in which a device remains within a 30-meter radius for at least 15 minutes, yielding approximately 175 million detected stops across the study period~\citep{liao2025effect}.
The mobility data are further processed to have the same structure as the US foot traffic data. \par

For the United States, we use Advan Weekly Patterns Plus data~\citep{advan2025weeklypatterns} throughout 2024.
This dataset provides weekly foot traffic aggregations for POIs derived from a panel of anonymized mobile device locations.
The data cover three metropolitan areas: New York (14.1 million POI-week records, 290,270 unique POIs), Washington DC (4.0 million records, 84,650 POIs), and Atlanta (4.8 million records, 102,181 POIs).
The dataset spans 53 weeks from January 2024 through early January 2025, capturing approximately 48 billion total visits across the three cities. Each record reports the number of visitors to a POI during a given week, along with the distribution of visitors' home CBGs, enabling computation of visitor diversity without individual-level tracking.

\subsubsection{Socioeconomic Data}
For Sweden, socioeconomic characteristics are obtained from Statistics Sweden (SCB) register-based statistics at the DeSO (Demografiska statistikområden) level for the year 2024~\citep{scb2024deso}.
DeSO zones (Demografiska statistikområden) are small statistical areas defined by Statistics Sweden with a median population of approximately 1,500 residents, designed to capture local demographic variation within municipalities.
Birth background data categorize the population into three groups: born in Sweden, born elsewhere in Europe, and born outside Europe. \par

For the United States, socioeconomic data come from the 2024 5-year American Community Survey (ACS) estimates at the census tract level (median population ~4,000)~\citep{uscensus2024acs}.
Birth background data distinguishes native-born from foreign-born residents.
To enable diversity calculations, we compute the proportion of each tract's population in each socioeconomic category.

\subsubsection{Transit Network Data}
For Sweden, transit network data were obtained from Samtrafiken AB via its publicly available API in the General Transit Feed Specification (GTFS) static format~\citep{liao2025effect}.
Public transit networks for the three US cities were obtained from their respective metropolitan transit agencies.
For New York, we combine GTFS feeds from the Metropolitan Transportation Authority (MTA), including the subway system, local and express bus routes, the Long Island Rail Road (LIRR), and Metro-North Railroad, capturing the full multimodal transit network serving the New York metropolitan area.
For Washington DC, we use GTFS data from the Washington Metropolitan Area Transit Authority (WMATA), encompassing the Metrorail heavy rail system and Metrobus network serving the District of Columbia, northern Virginia, and suburban Maryland.
For Atlanta, we obtain GTFS data from the Metropolitan Atlanta Rapid Transit Authority (MARTA), which operates both heavy rail and bus services in the Atlanta metropolitan area. \par

The GTFS data were up to date and processed to remove invalid route shapes and extract the geolocations of all transit stops.
OpenStreetMap road network data were obtained from Geofabrik to enable multimodal routing~\citep{geofabrik2026}.
The processed GTFS data and road network data serve as the basis for constructing transit-proximity spatial weights and for computing accessibility measures using \texttt{r5r}~\citep{r5r}, an open-source multimodal transport routing package.

\subsection{Data Preprocessing}\label{sec:data_proc}
\subsubsection{Mobility Data Processing (Sweden)}
Home locations are identified using the HoWDe (Home and Work Detection) algorithm~\citep{de2026howde}, which classifies stops based on temporal visitation patterns.
The algorithm identifies home locations from stops occurring during residential hours (before 6:00 and after 18:00) over a 14-day rolling window, applying density and frequency thresholds to distinguish consistent home patterns from transient visits.
This procedure achieves a 71.4\% home detection rate. \par

To ensure reliable home location quality, we link detected home coordinates to residential buildings from Overture Maps~\citep{overturemaps2024} using a 50-meter buffer to account for GPS positioning uncertainty.
Only devices with both a detected home location and a confirmed residential building match (85.2\% match rate) are retained for analysis.
Home locations are mapped to DeSO zones (Demographic Statistical Areas, median population of 1,500).
To correct for sampling bias in the mobile phone panel, we apply inverse probability weighting (IPW) based on the population size of each DeSO zone~\citep{liao2025effect,liao2025uncovering}. \par

Detected stays are matched to SafeGraph POIs using a tiered spatial assignment approach.
Stops falling inside a POI's building footprint polygon (available for 91\% of POIs) receive the highest confidence assignment; stops within 20 meters of a POI centroid are assigned with high confidence, accounting for GPS uncertainty; stops within 50 meters receive medium confidence; and stops within 100 meters receive low confidence assignments.
Stops within 100 meters of a device's home location are excluded to focus on out-of-home activities.
This tiered approach yields a 46.1\% match rate, producing approximately 80 million POI-linked visits that are aggregated to origin-zone-to-POI flows weighted by IPW, producing a dataset structurally comparable to the US foot traffic data.

\subsubsection{POI Category Processing and Harmonization}
To ensure cross-national comparability, we developed a unified POI categorization scheme that harmonizes venue classifications across Swedish and US datasets.
We mapped SafeGraph sub-category labels to 12 unified categories: food and dining, general retail, specialty retail, personal services, pharmacy and drug stores, entertainment and recreation, accommodation and travel, financial services, professional services, automotive, higher education, and civic and community venues.
We retained only categories achieving at least 50\% visitor home census block group coverage in the US foot traffic data, ensuring sufficient geographic information for reliable diversity estimation.
Several category groups were systematically excluded: healthcare providers (due to HIPAA privacy restrictions in US data), K-12 schools, and childcare facilities (available in the data for Sweden but not for the US), transit infrastructure (analyzed separately via GTFS data), and non-public venues such as manufacturing facilities, warehouses, and real estate offices.
To ensure reliable downstream calculations, we further filtered POIs to include only those with at least 50 total visits and visitors originating from at least 5 distinct residential zones, yielding final samples of approximately 200,000 POIs across the three US cities and 130,000 POIs across the Swedish cities.

\subsubsection{Transit Catchment Populations}\label{sec:catchment}
For each POI, we define its transit catchment as the set of residential zones reachable within 45 minutes by a combination of walking and public transit.
Transit accessibility is computed using the \texttt{r5r} package~\citep{r5r}, which implements the R5 routing engine to calculate realistic door-to-door travel times accounting for scheduled service, transfers, and walking segments.
We combine GTFS transit schedules with road network data to enable multimodal routing. \par

Travel times are computed for a typical weekday morning departure (8:00 AM) with a 15-minute time window to account for schedule variability, using the median travel time across departures within this window.
We set a maximum total trip duration of 45 minutes (including walking and waiting time), with a maximum walk time of 15 minutes at either end of the journey and an assumed walking speed of 4.5 km\/h.
Residential zones (census tracts in the US, DeSO zones in Sweden, represented by centroids) with median travel times at or below this threshold are included in the POI's catchment.
The socioeconomic composition of catchment populations forms the basis for quantifying potential social mixing at each POI.

\subsubsection{Cross-National Data Comparability}\label{sec:cross}

The Swedish and US analyses rely on different mobility data pipelines.
In Sweden, we process raw GPS trajectories through stop detection, home inference (HoWDe algorithm), and tiered spatial assignment to POIs, producing device-level visit flows that are subsequently aggregated.
In the US, the data provider has already performed these steps, delivering pre-aggregated weekly foot traffic records with visitor home-origin distributions per POI.
The two pipelines, therefore, differ in home location detection methods, POI matching procedures, and the degree of analyst control over processing decisions.
The spatial units used to characterize visitors' home neighborhoods also differ in definition (DeSO zones vs.\ census block groups), and the number of categories entering the birth-background entropy measure differs---two in Sweden versus five or more in the US---which affects absolute diversity values even after normalization. \par

We therefore treat the cross-national comparison as a context-dependent contrast between selected urban cases rather than a precise quantitative comparison.
Cross-national patterns are interpreted at the level of directional consistency and qualitative differences, while within-country variation across cities provides the more reliable basis for inference.

\subsection{Quantifying Social Mixing}\label{sec:social_mixing}
For each POI, we compute \textit{visitor diversity} as the entropy of the socioeconomic composition of its visitors' home neighborhoods.
Let $p_i$ denote the proportion of visitors originating from neighborhoods in socioeconomic category $i$.
The entropy-based diversity index is:

\begin{equation}
H = -\sum_{i=1}^{k} p_i \ln(p_i)
\label{eq:entropy}
\end{equation}

We normalize by maximum entropy ($\ln k$) to obtain values between 0 (all visitors from one category) and 1 (visitors evenly distributed across categories).
We compute separate indices for birth background (native-born vs. foreign-born).
For Sweden, the original three-category birth-background classification is collapsed into a two-group entropy measure: Sweden-born and born outside Europe, with Europe-born individuals excluded.
This is intended to focus the analysis on forms of migration-related diversity that are more likely to structure social boundaries and unequal exposure in everyday urban life, while avoiding the inclusion of Europe-born populations whose socioeconomic and cultural integration patterns may differ substantially from those of non-European migrants \citep{liao2025effect}. \par

Using the same metric (Equation~\ref{eq:entropy}), \textit{transit catchment diversity} measures the socioeconomic heterogeneity of populations reachable from each POI within 45 minutes by public transit (see Section~\ref{sec:catchment}), applied to the population-weighted socioeconomic distribution of reachable census zones. \par

\textit{Residential diversity} captures the socioeconomic composition of the POI's immediate neighborhood, computed as the entropy of the tract/DeSO zone in which the POI is located.
This controls for the baseline expectation that POIs in diverse neighborhoods attract diverse visitors solely because of proximity.

\subsection{Characterizing Locational Context of POIs}\label{sec:spatial}
Three predictors are selected to capture distinct and theoretically motivated dimensions of a POI's locational context.
Distance to city center indexes a POI's position along the urban centrality gradient, which is associated with higher visitor volumes, greater commercial agglomeration, and denser transit service \citep{ewing2010travel, cervero1997travel}.
POI density within 500 m captures local commercial clustering, which may, in turn, generate visitor diversity through agglomeration externalities and pedestrian spillovers between neighboring establishments \citep{hidalgo2020amenity}.
Distance to the nearest transit stop measures the physical accessibility of the POI to transit infrastructure, reflecting how directly a venue is embedded within the transit network.
Together, these three variables characterize the spatial opportunity structure surrounding each POI --- centrality, local commercial intensity, and transit proximity --- without overlapping with the diversity measures that define the dependent variable. \par

\textbf{Distance to city center}.
For each POI, we compute the geodesic distance to the city center (in kilometers) using the Haversine formula.
City center coordinates are defined as the central business district location for each city (e.g., Times Square for New York, Central Station for Stockholm).
This variable captures whether transit--diversity effects concentrate in central urban areas or extend to peripheral locations. \par

\textbf{POI density}.
We measure local commercial intensity as the count of other POIs within a 500-meter radius of each focal POI.
This is computed efficiently using a KD-tree spatial index on projected coordinates.
Higher POI density indicates clustering of commercial activity, which may facilitate or reflect pedestrian flows from nearby transit infrastructure. \par

\textbf{Transit stop proximity}.
We compute the Euclidean distance (in meters) from each POI to its nearest transit stop, using GTFS (General Transit Feed Specification) data for each city.
This variable directly measures physical accessibility to transit infrastructure. \par

\subsection{Describing Spatial Structure of Social Mixing}\label{sec:desc_methods}
We assess spatial clustering in visitor diversity using global and local measures of spatial autocorrelation.
For each city, POIs are first projected to the appropriate UTM coordinate system to ensure accurate distance calculations.
We construct a $k$-nearest neighbors (KNN) spatial weights matrix with $k = 10$, connecting each POI to its 10 nearest neighbors based on Euclidean distance in projected coordinates.
The weight matrix is row-standardized so that the weights for each POI sum to unity, ensuring that the spatial lag represents a weighted average of neighboring values. \par

We compute Moran's I statistic to quantify the overall degree of spatial clustering in visitor diversity:
\begin{equation}
    I = \frac{n}{\sum_i \sum_j w_{ij}} \frac{\sum_i \sum_j w_{ij}(y_i - \bar{y})(y_j - \bar{y})}{\sum_i (y_i - \bar{y})^2}
\end{equation}
where $n$ is the number of POIs, $w_{ij}$ are elements of the row-standardized spatial weights matrix, $y_i$ is visitor diversity (normalized entropy) at POI $i$, and $\bar{y}$ is the mean visitor diversity.
Values of $I > 0$ indicate positive spatial autocorrelation (similar values cluster together), while $I < 0$ indicates negative autocorrelation (dissimilar values are neighbors). \par

To identify statistically significant local clusters (Local Indicators of Spatial Association, LISA), we compute Local Moran's I for each POI \citep{anselin1995local}:
\begin{equation}
    I_i = \frac{(y_i - \bar{y})}{\sigma^2_y} \sum_j w_{ij}(y_j - \bar{y})
\end{equation}
where $\sigma^2_y$ is the variance of $y$. \par

Statistical inference is based on a conditional permutation approach: for each POI, we hold its value fixed while randomly permuting the values at all other locations, repeating this process 100 times to construct a reference distribution under the null hypothesis of spatial randomness.
A POI is classified as belonging to a significant cluster if its pseudo $p$-value (the proportion of permuted statistics as extreme as the observed) is below $\alpha = 0.05$. \par

Significant POIs are assigned to one of four quadrants based on the relationship between the focal POI's value and its neighbors' values:
\begin{itemize}
    \item \textbf{High-High (HH)}: High-diversity POIs surrounded by high-diversity neighbors (diversity hotspots)
    \item \textbf{Low-Low (LL)}: Low-diversity POIs surrounded by low-diversity neighbors (diversity coldspots)
    \item \textbf{High-Low (HL)}: High-diversity POIs surrounded by low-diversity neighbors (spatial outliers)
    \item \textbf{Low-High (LH)}: Low-diversity POIs surrounded by high-diversity neighbors (spatial outliers)
\end{itemize}

We acknowledge that the uneven spatial distribution of POIs introduces scale heterogeneity into the spatial autocorrelation measures.
In dense central areas, the 10 nearest neighbors span a smaller geographic radius than in peripheral areas, meaning Moran's $I$ and LISA statistics capture spatial association at varying resolutions across the study area.
The KNN specification with row standardization ensures that each POI's spatial lag is computed from exactly $k = 10$ neighbors regardless of local density, avoiding the variable neighborhood sizes that would arise with fixed-distance weights.
However, alternative specifications such as distance-band or density-adjusted weights could be explored in future work to assess sensitivity to this issue.

\subsection{Models for Transit Catchment Effects on Visitor Diversity}

To examine whether transit catchment diversity predicts visitor diversity at POIs, we employ two complementary regression approaches: ordinary least squares (OLS) as a baseline and spatial lag models (SLM) to account for spatial dependence.

\subsubsection{OLS Baseline Model}
We first estimate a standard OLS regression for each city and diversity dimension (birth background):
\begin{equation}
    V_i = \alpha + \beta_1 R_i + \beta_2 T_i + \boldsymbol{\gamma}' \mathbf{D}_i + \varepsilon_i
\end{equation}
where $V_i$ is the normalized visitor diversity at POI $i$, $R_i$ is the residential diversity of the census tract containing the POI, $T_i$ is the transit catchment diversity (the population-weighted diversity of areas reachable within 45 minutes by public transit), $\mathbf{D}_i$ is a vector of dummy variables for POI category (with the five most common categories in each city plus an ``other'' category, one category omitted as reference), and $\varepsilon_i$ is the error term.
The coefficient $\beta_2$ captures the independent contribution of transit accessibility to visitor diversity, controlling for local residential composition and venue type.

\subsubsection{Stepwise Robustness Check}
We assess the robustness of the transit catchment coefficient by estimating a sequence of nested OLS specifications for each city and diversity dimension.
Unlike the baseline model, which controls only for residential diversity and POI category, the robustness sequence progressively adds spatial controls that may confound the relationship between transit-accessible diversity and visitor diversity. \par

The estimated sequence is:
\begin{align}
    V_i &= \alpha + \beta_1 R_i + \beta_2 T_i
        + \boldsymbol{\gamma}' \mathbf{D}_i + \varepsilon_i \label{eq:robust_base} \\
    V_i &= \alpha + \beta_1 R_i + \beta_2 T_i + \beta_3 G_i
        + \boldsymbol{\gamma}' \mathbf{D}_i + \varepsilon_i \label{eq:robust_geo} \\
    V_i &= \alpha + \beta_1 R_i + \beta_2 T_i + \beta_3 G_i
        + \beta_4 C_i + \boldsymbol{\gamma}' \mathbf{D}_i + \varepsilon_i \label{eq:robust_center} \\
    V_i &= \alpha + \beta_1 R_i + \beta_2 T_i + \beta_3 G_i
        + \beta_4 C_i + \beta_5 A_i
        + \boldsymbol{\gamma}' \mathbf{D}_i + \varepsilon_i \label{eq:robust_full}
\end{align}
where $G_i$ is geographic catchment diversity, measured as the population-weighted diversity within a 1.5 km Euclidean buffer around the POI; $C_i$ is distance to the city center in kilometers; and $A_i$ is the number of POIs within 500 meters, used as a local agglomeration proxy.
The category controls $\mathbf{D}_i$ are constructed as in the baseline model, using the five most common POI categories in each city plus an ``other'' category, with one category omitted as the reference group. \par

This stepwise design tests whether the estimated transit catchment association is robust to three alternative spatial explanations (see Section \ref{sec:spatial}).
The geographic catchment term accounts for diversity in the nearby residential environment independent of transit routing.
Distance to the city center captures centrality effects, since central POIs may be both more transit-accessible and more socially mixed.
Local POI density captures agglomeration, suggesting that dense commercial areas attract more heterogeneous visitors regardless of transit catchment composition. \par

For each specification, we compare the magnitude, sign, standard error, and statistical significance of $\beta_2$, along with changes in model fit measured by $R^2$, adjusted $R^2$, and AIC.
If $\beta_2$ remains positive and statistically significant after adding $G_i$, $C_i$, and $A_i$, this provides evidence that transit catchment diversity is not merely proxying for nearby residential diversity, urban centrality, or local venue density.

\subsubsection{Spatial Lag Model}\label{sec:slm}
OLS estimates may be biased if visitor diversity exhibits spatial dependence, i.e., if diversity at one POI is correlated with that at nearby POIs.
To address this, we estimate a spatial lag model (SLM):
\begin{equation}
    V_i = \rho \sum_{j} w_{ij} V_j + \alpha + \beta_1 R_i + \beta_2 T_i + \boldsymbol{\gamma}' \mathbf{D}_i + \varepsilon_i
\end{equation}
where $\rho$ is the spatial autoregressive parameter capturing the degree of spatial spillover, $w_{ij}$ are elements of a row-standardized spatial weights matrix $\mathbf{W}$ defining the neighborhood structure, and $\mathbf{D}_i$ represents venue category fixed effects as defined above.
A significant positive $\rho$ indicates that visitor diversity at a POI is influenced by visitor diversity at neighboring POIs.
The SLM is estimated using the generalized method of moments (GMM) \citep{kelejian1999generalized}. \par

For valid interpretation, the spatial lag model requires stationarity: $|\rho| < 1$.
When this condition holds, the spatial multiplier $(I - \rho W)^{-1}$ converges and total effects can be meaningfully decomposed into direct and indirect components.
Results where $|\rho| \geq 1$ indicate non-stationarity, suggesting model misspecification (e.g., omitted spatially-structured variables) rather than genuine spillover effects; such cases are flagged and excluded from substantive interpretation. \par

We construct two alternative spatial weights matrices to test different spillover mechanisms:
W1 indicates the distance-decay weights with a cutoff, where POIs are connected based on Euclidean distance with inverse-distance weighting and a maximum threshold:
\begin{equation}
    w_{ij}^{(1)} = \begin{cases}
        d_{ij}^{-\alpha} & \text{if } d_{ij} \leq d_{\max} \text{ and } i \neq j \\
        0 & \text{otherwise}
    \end{cases}
\end{equation}
where $d_{ij}$ is the distance between POIs $i$ and $j$ in meters, $\alpha = 1$ is the decay exponent, and $d_{\max} = 800$m is the distance cutoff.
This specification captures pedestrian-scale spatial spillover: nearby POIs receive higher weights than distant ones, and POIs beyond typical walking distance receive zero weight.
Unlike $k$-nearest neighbors, distance-decay weights distinguish between neighbors at different distances (e.g., 50m vs. 500m) rather than treating all neighbors equally. \par

W2 indicates shared transit stop weights, where POIs are connected if they share the same nearest transit stop.
Each POI is first assigned to its nearest transit stop within a 400-meter radius; two POIs are then considered neighbors if and only if they are assigned to the same stop:
\begin{equation}
    w_{ij}^{(2)} = \begin{cases}
        1 & \text{if } s^*(i) = s^*(j) \text{ and } d_{i,s^*(i)} \leq 400\text{m} \text{ and } d_{j,s^*(j)} \leq 400\text{m} \\
        0 & \text{otherwise}
    \end{cases}
\end{equation}
where $s^*(i) = \arg\min_s d_{is}$ is the nearest transit stop to POI $i$, and $d_{is}$ is the distance from POI $i$ to stop $s$.
This specification captures transit-mediated connectivity: POIs clustered around the same transit node are assumed to draw from overlapping visitor pools.
Both matrices are row-standardized such that $\sum_j w_{ij} = 1$ for all $i$ with at least one neighbor. \par

This operationalization of W2 captures one dimension of transit-mediated connectivity but does not account for service frequency, network connectivity, or modal hierarchy, any of which could make a slightly more distant stop functionally more relevant than the nearest one.
Conversely, adjacent stops on the same line may provide nearly identical access but are treated as distinct nodes.
We retain this specification for its parsimony and reproducibility across cities, but acknowledge that alternative formulations incorporating route diversity or stop-level service attributes could yield different patterns of transit-mediated spatial dependence. \par

We compare spatial lag models with different weight specifications to test whether transit-specific connectivity (W2) better explains spatial dependence than general proximity (W1).
A larger $\rho$ under W2 than W1 would suggest that transit infrastructure specifically channels diversity spillovers between POIs, supporting the hypothesis that transit nodes act as spatial bridges for social mixing.

\subsection{Geographically Weighted Regression}\label{sec:gwr}

While OLS and spatial lag models estimate global relationships, they assume that the effect of transit catchment diversity on visitor diversity is spatially constant.
To examine whether this relationship varies geographically—and to identify locations where transit infrastructure most strongly shapes social mixing—we employ Geographically Weighted Regression (GWR). \par

GWR extends the global regression framework by allowing coefficients to vary continuously across space.
For each POI $i$, we estimate a local model:
\begin{equation}
    V_i = \beta_0(u_i, v_i) + \beta_1(u_i, v_i) R_i + \beta_2(u_i, v_i) T_i + \varepsilon_i
\end{equation}
where $(u_i, v_i)$ are the projected coordinates of POI $i$, $V_i$ is visitor diversity, $R_i$ is residential diversity, $T_i$ is transit catchment diversity, and $\beta_k(u_i, v_i)$ are location-specific coefficients.
Each local regression is estimated using a spatial kernel that assigns higher weights to nearby observations. \par

We use an adaptive bisquare kernel with bandwidth defined by the distance to the $k$th nearest neighbor.
The optimal bandwidth is selected by minimizing the corrected Akaike Information Criterion (AICc):
\begin{equation}
    \text{AICc} = 2n \ln(\hat{\sigma}) + n \ln(2\pi) + n \left( \frac{n + \text{tr}(\mathbf{S})}{n - 2 - \text{tr}(\mathbf{S})} \right)
\end{equation}
where $n$ is the sample size, $\hat{\sigma}$ is the residual standard deviation, and $\text{tr}(\mathbf{S})$ is the trace of the hat matrix, representing the effective number of parameters.
To ensure numerical stability in local estimation, we impose a minimum bandwidth of $\max(50, 0.01n)$ neighbors. \par

Prior to estimation, POI coordinates are projected from WGS84 to the appropriate UTM zone for each city to ensure accurate distance calculations.
To address duplicate or near-duplicate coordinates (e.g., multiple POIs within the same building), we add small random jitter (approximately 1 meter standard deviation).
Independent variables are standardized to improve numerical convergence, and coefficients are subsequently transformed back to their original scale for interpretation. \par

Statistical inference in GWR requires adjustment for multiple, spatially dependent hypothesis tests.
We apply the da Silva–Fotheringham correction \citep{da2016multiple}, which accounts for the effective number of independent tests implied by spatial smoothing.
The effective number of tests is given by
\begin{equation}
    n_{\text{eff}} = \frac{n}{\text{tr}(\mathbf{S})},
\end{equation}
yielding a corrected significance threshold below.

\begin{equation}
    \alpha_{\text{corrected}} = \alpha \cdot \frac{\text{tr}(\mathbf{S})}{n} = \frac{\alpha}{n_{\text{eff}}}.
\end{equation}

Local $t$-statistics are computed as
\begin{equation}
    t_i = \frac{\hat{\beta}_k(u_i, v_i)}{\text{SE}_k(u_i, v_i)},
\end{equation}
with $p$-values based on a two-tailed $t$-test using effective degrees of freedom $df = n - \text{tr}(\mathbf{S})$.
Coefficients are considered significant when $p_i < \alpha_{\text{corrected}}$.
We interpret local significance cautiously and primarily use it to identify broad spatial patterns rather than strict pointwise inference. \par

To characterize spatial heterogeneity, we summarize the distribution of the local catchment coefficient $\hat{\beta}_2(u_i, v_i)$ using its mean, standard deviation, interquartile range, and coefficient of variation:
\begin{equation}
    \text{CV} = \frac{\text{SD}(\hat{\beta}_2)}{\left| \overline{\hat{\beta}_2} \right|}.
\end{equation}
We further identify \textit{transit--diversity hotspots} as locations where the catchment coefficient is both positive and statistically significant under the corrected threshold, and \textit{coldspots} as locations with significant negative effects.
This classification highlights where transit accessibility most strongly—and least strongly—predicts visitor diversity. \par

GWR models are estimated separately for each combination of city and diversity dimension (birth background).
We evaluate both POI-category-stratified GWR models and the GWR models using all POIs.
All POIs meeting data completeness requirements are included, and strata with fewer than 100 observations are excluded due to insufficient sample size for stable local estimation. \par

We note that our adaptive kernel specification addresses the well-known limitation of fixed-bandwidth GWR, where a single distance threshold cannot accommodate variation in observation density across the study area.
A further extension, Multiscale GWR (MGWR), allows each covariate to have its own bandwidth, thereby revealing differences in the spatial scale at which individual predictors operate \citep{fotheringham2017multiscale}.
We use standard adaptive GWR here because our primary interest lies in mapping the spatial heterogeneity of the transit catchment coefficient rather than comparing operational scales across covariates, and because MGWR's iterative back-fitting procedure is computationally prohibitive at the sample sizes in this study (exceeding 100,000 POIs in several cities).

\subsection{Characterizing Transit--Diversity Hotspots}\label{sec:gwr_hotspots}

To understand the spatial heterogeneity of transit--diversity effects identified through GWR, we examine the locational and contextual characteristics that distinguish hotspots---POIs where transit catchment diversity significantly predicts visitor diversity---from POIs with non-significant effects.
Hotspot status is derived from the pooled GWR specification model, which jointly fits all POIs without stratifying by venue category, so the results characterize heterogeneity across the full set of POIs rather than within specific venue types. \par

This analysis addresses the question: \textit{where} do transit-mediated mixing effects concentrate, and what features characterize these locations?
We adopt pre-defined spatial contextual variables (see Section \ref{sec:spatial}). \par

To jointly model the predictors of hotspot status, we estimate a logistic regression:
\begin{equation}
    \text{logit}\left( P(\text{Hotspot}_i = 1) \right) = \mathbf{X}_i \boldsymbol{\beta}
\end{equation}
where $\text{Hotspot}_i$ is a binary indicator for whether POI $i$ has a significant positive transit catchment coefficient in the GWR model, $\mathbf{X}_i$ is a vector of standardized predictors, and $\boldsymbol{\beta}$ are the regression coefficients. \par

The predictors include the spatial context variables defined above: distance to city center (km, standardized), POI density within 500m (standardized), and distance to nearest transit stop (m, standardized).
We deliberately exclude transit catchment diversity and residential diversity from this model, as these variables are already included in the GWR specification that defines hotspot status. \par

We estimate separate models for Swedish and US cities to allow for context-specific effects reflecting differences in urban form and transit infrastructure.
Models are estimated using maximum likelihood estimation for binary response models.
We report odds ratios (OR) with 95\% confidence intervals, computed as $\exp(\hat{\beta} \pm 1.96 \times \text{SE})$.
An odds ratio greater than 1 indicates that a one-standard-deviation increase in the predictor is associated with higher odds of being a hotspot.
We assess model performance using the area under the receiver operating characteristic (ROC) curve (ROC-AUC), computed from predicted probabilities, along with classification accuracy at a threshold of 0.5.

\section*{Conflict of Interest Statement}
The author declares that the research was conducted in the absence of any commercial or financial relationships that could be construed as a potential conflict of interest.

\section*{Author Contributions}
YL conceived and designed the study, performed data collection and analysis, interpreted the results, and wrote and revised the manuscript.

\section*{Funding}
This research is funded by the Swedish Research Council (Project Number 2022-06215).

\section*{Acknowledgments}
The author acknowledges Jorge Gil for providing the mobile phone application data for Sweden.

\section*{Data Availability Statement}
The datasets used in this study have varying access conditions.
Publicly available data include 1) Swedish socioeconomic data (DeSO zones), available from Statistics Sweden (SCB) \citep{scb2024deso}, 2) US Census data (American Community Survey 5-year estimates), available via the Census Bureau API \citep{uscensus2024acs}, 3) Swedish transit schedules (GTFS), available from Trafiklab at \url{https://www.trafiklab.se/}, and 4) US transit schedules (GTFS) are available from the cities' respective data portals.
Commercially licensed data include 1) point of interest data from SafeGraph Global Places \citep{safegraph2022globalplaces} and 2) US foot traffic data from Advan Weekly Patterns Plus \citep{advan2025weeklypatterns}.
Restricted data include Swedish mobility data (anonymized GPS trajectories) contain sensitive location information and cannot be shared publicly, due to licensing and privacy considerations under the European General Data Protection Regulation.
Analysis code is available at \url{https://github.com/MobiSegInsights/geo-social-mixing}.

\section*{Generative AI Statement}
During the preparation of this manuscript, the author used ChatGPT (OpenAI, GPT-5 series) and Claude (Anthropic, Opus 4 series) to assist with language editing, restructuring, and refinement of text.
During the preparation of this manuscript, the author used Claude Code (Anthropic, Sonnet 4.5 and Opus 4 series) to assist with coding, data analysis, and manuscript editing.
All AI-assisted content was critically evaluated, verified against original sources, and substantially revised by the author, who takes full responsibility for the final content.

\bibliographystyle{unsrtnat}
\bibliography{references}

\newpage
\appendix
\renewcommand{\thefigure}{S.\arabic{figure}}
\renewcommand{\thetable}{S.\arabic{table}}
\setcounter{figure}{0}
\setcounter{table}{0}

\section{Supplementary Tables}

\begin{table}[htbp]
  \centering
  \caption{Spatial clustering of visitor diversity by birth background across Swedish and US cities. Moran's I measures global spatial autocorrelation; HH\% and LL\% indicate the percentage of POIs in High-High and Low-Low LISA clusters, respectively. All Moran's I values are significant at $p < 0.001$.}
  \label{tab:spatial_clustering_full}
  \begin{tabular}{llrccc}
  \toprule
  \textbf{Country} & \textbf{City} & \textbf{N POIs} & \textbf{Moran's I} & \textbf{HH\%} & \textbf{LL\%} \\
  \midrule
   & Gothenburg   & 16,685  & 0.237 & 9.3  & 14.5 \\
   & Helsingborg  & 5,275   & 0.161 & 11.4 & 10.7 \\
   & Linköping    & 5,311   & 0.182 & 7.4  & 10.1 \\
  Sweden & Lund         & 4,824   & 0.081 & 5.0  & 5.7  \\
   & Malmö        & 12,610  & 0.197 & 14.3 & 12.8 \\
   & Stockholm    & 35,842  & 0.127 & 6.0  & 5.6  \\
   & Uppsala      & 7,933   & 0.167 & 10.7 & 10.1 \\
   & Västerås     & 5,263   & 0.142 & 9.4  & 9.7  \\
   & Örebro       & 5,407   & 0.131 & 5.3  & 7.1  \\
  \midrule
   & New York       & 148,413 & 0.833 & 38.6 & 17.7 \\
  US & Atlanta        & 34,354  & 0.775 & 26.2 & 28.9 \\
   & Washington DC  & 24,599  & 0.667 & 31.2 & 16.3 \\
  \bottomrule
  \end{tabular}
\end{table}

\begin{table}[htbp]
  \centering
  \caption{Comparison of residential, visitor, and transit catchment diversity by birth background.
  $^{***}p<0.001$, $^{**}p<0.01$, $^{*}p<0.05$, $^{ns}$ not significant.}
  \label{tab:merged_summary}
  \begin{tabular}{lrrrrrrr}
  \toprule
  \textbf{City} & \textbf{N} & \textbf{Res.} & \textbf{Vis.} & \textbf{Cat.} &
  \textbf{Vis.-Res.} & \textbf{Cat.-Res.} & \textbf{Diff.-in-Diff.} \\
  \midrule
  New York      & 148,413 & 0.825 & 0.873 & 0.887 & +0.048$^{***}$ & +0.062$^{***}$ & -0.014$^{***}$ \\
  Washington DC & 24,599  & 0.826 & 0.840 & 0.872 & +0.014$^{ns}$  & +0.046$^{***}$ & -0.032$^{***}$ \\
  Atlanta       & 34,354  & 0.592 & 0.607 & 0.626 & +0.015$^{***}$ & +0.034$^{***}$ & -0.019$^{***}$ \\
  \midrule
  Gothenburg    & 16,685  & 0.575 & 0.607 & 0.718 & +0.032$^{***}$ & +0.143$^{***}$ & -0.111$^{***}$ \\
  Helsingborg   & 5,275   & 0.614 & 0.598 & 0.743 & -0.015$^{***}$ & +0.130$^{***}$ & -0.145$^{***}$ \\
  Linköping     & 5,311   & 0.539 & 0.530 & 0.634 & -0.008$^{*}$   & +0.095$^{***}$ & -0.103$^{***}$ \\
  Lund          & 4,824   & 0.561 & 0.575 & 0.645 & +0.014$^{***}$ & +0.084$^{***}$ & -0.070$^{***}$ \\
  Malmö         & 12,610  & 0.673 & 0.652 & 0.806 & -0.021$^{***}$ & +0.133$^{***}$ & -0.154$^{***}$ \\
  Stockholm     & 35,842  & 0.454 & 0.583 & 0.619 & +0.129$^{***}$ & +0.166$^{***}$ & -0.037$^{***}$ \\
  Uppsala       & 7,933   & 0.539 & 0.592 & 0.693 & +0.053$^{***}$ & +0.155$^{***}$ & -0.101$^{***}$ \\
  Västerås      & 5,263   & 0.601 & 0.614 & 0.733 & +0.013$^{***}$ & +0.132$^{***}$ & -0.118$^{***}$ \\
  Örebro        & 5,407   & 0.480 & 0.507 & 0.609 & +0.028$^{***}$ & +0.129$^{***}$ & -0.102$^{***}$ \\
  \bottomrule
  \end{tabular}
\end{table}

\begin{table}[htbp]
  \centering
  \caption{GWR summary statistics of spatial heterogeneity in the transit catchment diversity's coefficient across cities.
  Values report distributional characteristics of local coefficients and model fit.
  \%+ denotes the percentage of locations with positive coefficients; \%Hot and \%Cold indicate the shares of statistically significant positive and negative coefficients, respectively.}
  \label{tab:gwr_summary}
  \begin{tabular}{lrrrrrr}
  \toprule
  \textbf{City} & \textbf{Mean} & \textbf{SD} & \textbf{\% Positive} & \textbf{\%Hot} & \textbf{\%Cold} & \textbf{$R^2$} \\
  \midrule
    Stockholm          & .187 & 1.350   & 53.9 & 0.5  & 1.2 & .118 \\
    Göteborg           & .225 & 1.685   & 51.1 & 0.8  & 0.4 & .225 \\
    Malmö              & -.123 & 1.731  & 47.7 & 0.3  & 1.1 & .182 \\
    Uppsala            & -.124 & 2.960  & 52.5 & 0.3  & 2.6 & .185 \\
    Helsingborg        & -.117 & 1.358  & 29.3 & 6.1  & 8.8 & .128 \\
    Lund               & .081 & 1.285   & 55.7 & 0.1  & 0.0 & .079 \\
    Västerås           & 1.599 & 8.897  & 55.9 & 1.7  & 1.2 & .152 \\
    Örebro             & -.275 & 1.729  & 45.4 & 1.0  & 0.0 & .141 \\
    Linköping          & .213 & 2.090   & 55.5 & 1.0  & 0.1 & .186 \\
    New York (US)      & .150 & .834    & 64.8 & 32.5 & 8.1 & .794 \\
    Atlanta (US)       & .118 & .317    & 68.9 & 12.1 & 1.5 & .766 \\
    Washington DC (US) & .915 & 133.594 & 62.6 & 7.5  & 2.1 & .673 \\
  \bottomrule
  \end{tabular}
\end{table}

\begin{table}[htbp]
  \centering
  \caption{Logistic regression predicting transit-mixing hotspots.
  Odds ratios (OR) are reported with 95\% confidence intervals.
  Predictors are standardized (1 unit = 1 SD).
  $^{*}p<0.05$, $^{**}p<0.01$, $^{***}p<0.001$.
  The hotspots are derived from the GWR results for all POIs pooled within each country's cities.}
  \label{tab:logit_hotspot}
  \begin{tabular}{lrr}
  \toprule
  \textbf{Variable} & \textbf{Sweden (9)} & \textbf{US (3)} \\
  \midrule
  Distance to center (km) & .111 [.081, .151]$^{***}$ & 1.171 [1.158, 1.185]$^{***}$ \\
  POI density (500 m)      & .705 [.646, .769]$^{***}$ & .956 [.945, .967]$^{***}$ \\
  Transit proximity (m)   & 3.934 [.297, 52.136]      & 1.301 [1.284, 1.318]$^{***}$ \\
  \addlinespace[2pt]
  Pseudo $R^2$            & .026 & .023 \\
  ROC-AUC                 & .673 & .582 \\
  Accuracy                & .991 & .721 \\
  Sensitivity             & .000 & .044 \\
  Specificity             & 1.000 & .983 \\
  N (K)                   & 97.8 & 194.3 \\
  Hotspots (\%)           & .9 & 27.9 \\
  \bottomrule
  \end{tabular}
\end{table}

\end{document}